\documentclass[lettersize,journal]{IEEEtran}
\usepackage{amsmath,amssymb,amsfonts}
\usepackage{algorithmic}
\usepackage{algorithm}
\usepackage{array}
\usepackage{textcomp}
\usepackage{url}
\usepackage{verbatim}
\usepackage{graphicx}
\usepackage{cite}
\usepackage{subcaption}
\usepackage{xcolor}
\usepackage{caption}
\usepackage{ulem}
\def\BibTeX{{\rm B\kern-.05em{\sc i\kern-.025em b}\kern-.08em
    T\kern-.1667em\lower.7ex\hbox{E}\kern-.125emX}}
\begin{document}
\title{Enhancing Next-Generation Urban Connectivity: Is the Integrated HAPS-Terrestrial Network a Solution?}
\author{Afsoon~Alidadi~Shamsabadi,~\IEEEmembership{Member~,~IEEE,}~Animesh~Yadav,~\IEEEmembership{Senior~Member,~IEEE,} Halim~Yanikomeroglu,~\IEEEmembership{Fellow,~IEEE}
\thanks{Afsoon Alidadi Shamsabadi and Halim Yanikomeroglu are with Non-Terrestrial Networks Lab., Department of Systems and Computer Engineering, Carleton University, Ottawa, ON, K1S 5B6, Canada (e-mail: \{afsoonalidadishamsa, halim\}@sce.carleton.ca). Animesh Yadav is with the School of Electrical Engineering and Computer Science, Ohio University, Athens, OH, 45701, USA (e-mail: yadava@ohio.edu).}
}
\maketitle
\vspace{-1cm}
\begin{abstract}
Located in the stratospheric layer of Earth's atmosphere, high altitude platform station (HAPS) is a promising network infrastructure, which can bring significant advantages to sixth-generation (6G) and beyond wireless communications systems by forming vertical heterogeneous networks (vHetNets). However, if not dealt with properly, integrated networks suffer from several performance challenges compared to standalone networks. In harmonized spectrum integrated networks, where different tiers share the same frequency spectrum, interference is an important challenge to be addressed. This work focuses on an integrated HAPS-terrestrial network, serving users in an overlapped urban geographic area, and formulates a fairness optimization problem, aiming to maximize the minimum spectral efficiency (SE) of the network. Due to the highly nonconvex nature of the formulated problem, we develop a rapid converging iterative algorithm that designs the multiple-input multiple-output (MIMO) beamforming weights and the user association scheme such that the propagated inter- and intra-tier interference is managed. Simulation results demonstrate the proposed algorithm's superiority over standalone terrestrial networks and scenario where only the beamforming weights are optimized.
\end{abstract}

\begin{IEEEkeywords}
HAPS, vHetNets, interference, user association, MIMO beamforming, spectral efficiency
\end{IEEEkeywords}

\section{Introduction}
The rapid increase in mobile traffic and the emergence of novel use cases have led to a pressing need for network architectures that can cope with the forthcoming network trend in sixth-generation (6G) and beyond wireless communications systems. In this context, non-terrestrial networks (NTNs) have emerged as a promising architecture for incorporating powerful technologies into communication networks\cite{NTN-6G}. Particularly, High altitude platform stations (HAPS) are potential network platforms for supporting mobile users and operators in facilitating various use cases such as computing, sensing, positioning, and more. HAPS advantages such as lower latency than satellites, and wider coverage area than terrestrial base stations, makes it a promising complementary platform to existing wireless communications infrastructure\cite{HAPSSurvey}\cite{HAPS-2001}. HAPS can play the role of a super macro base station \cite{HAPSbasestation}, and be integrated with conventional terrestrial base stations to form vertical heterogeneous networks (vHetNets). Clearly, vHetNets improve the coverage of telecommunications networks due to HAPS and can significantly improve the network capacity, especially in hotspot urban areas where conventional terrestrial networks cannot meet the increasing traffic demand.

However, harmonized spectrum integrated networks, where different tiers share the same frequency band, are prone to performance degradation due to inter-tier interference propagation\cite{Spectrum Sharing}. To address this issue, efficient interference management and coordination techniques must be implemented in such multi-tier networks. Several techniques have been proposed to mitigate the interference challenge\cite{HAPSIM_0,HAPSIM-01,HAPSIM_2,HAPSIM_4,UA,Alouini,myletter}. In \cite{HAPSIM_0}, the authors proposed a HAPS-based null broadening approach for interference suppression in an integrated network. The authors in \cite{HAPSIM-01} maximized the HAPS coverage by decreasing the interference in the coexistence with terrestrial networks by optimizing the beamwidth and beam direction of HAPS antenna. Beamforming was employed in \cite{HAPSIM_2} to mitigate interference from HAPS to terrestrial cells. In \cite{HAPSIM_4}, the authors proposed an interference coordination method for an integrated HAPS-terrestrial network considering the traffic load distribution. In \cite{UA}, the authors developed a user association scheme for an integrated HAPS-terrestrial network based on the deep Q-learning (DQL) approach considering delayed channel state information (CSI). The authors in \cite{Alouini} developed an alternative user association and beamforming algorithm in an integrated HAPS-terrestrial-satellite network, where HAPS plays the role of the relay in satellite-UE connection. In \cite{myletter}, we developed an iterative algorithm to design the subcarrier and transmit power allocation to user equipment (UEs) to handle the interference in vHetNets.

Compared with \cite{myletter}, this work differs in four ways. First, we consider HAPS and terrestrial macro base stations (MBSs) are equipped with multiple-input multiple-output (MIMO) antennas, and hence, deploy beamforming scheme to form narrow high-gain beams toward served UEs. Secondly, we assume all the UEs are using the same time-frequency resource. Therefore, a proper design of the beamforming weights is required to generate user-specific beams toward the UEs and to control the allocated power to each UE, which is an important factor in the interference management of vHetNets\cite{myletter}. Thirdly, we also design a user association scheme as it plays an important role in the interference mitigation of the vHetNets, mainly due to the high transmit power and line-of-sight (LoS) connection between HAPS and UEs. Lastly, we employed a realistic channel model. Particularly, between the HAPS and UEs, a three-dimensional (3D) Rician fading channel with a dominant LoS connection and non-LoS (NLoS) parts is employed \cite{HAPS-MIMO} and shadowing factor is included in the channel between the MBSs and UEs.

The main contributions of this paper can be summarized as follows:
\begin{itemize}
    \item We formulated a max-min fairness (MMF) problem that maximizes the network's minimum spectral efficiency (SE) and designs the joint user association scheme and beamforming weights.
    \item Since the formulated problem is a mixed-integer non-linear program (MINLP), we employed reformulation linearization techniques (RLT) to recast the problem to an approximate convex format. Accordingly, we proposed a rapid converging algorithm based on successive convex approximation (SCA) to solve the approximate problem.
    \item We provided simulation results to compare the performance of vHetNets, under proposed algorithm, with baseline schemes and standalone terrestrial networks. The results validated the performance improvement, achieved by integrating HAPS with terrestrial networks, compared to the standalone terrestrial networks.
\end{itemize}
	
The remainder of the paper is organized as follows.  Section \ref{Sec:model} presents the system model and formulates the optimization problem, Section \ref{Sec:Algorithm} details the problem relaxation and explains the proposed algorithm, Section \ref{Sec:Results} presents and discusses the obtained numerical results, and finally, Section \ref{Sec:Conclusion}  concludes the paper.
\vspace{-0.3cm}
\section{System Model and Problem Formulation} \label{Sec:model}
\begin{figure}[t]
    \centering
    \captionsetup{justification=centering}
    \includegraphics[width=0.9\linewidth]{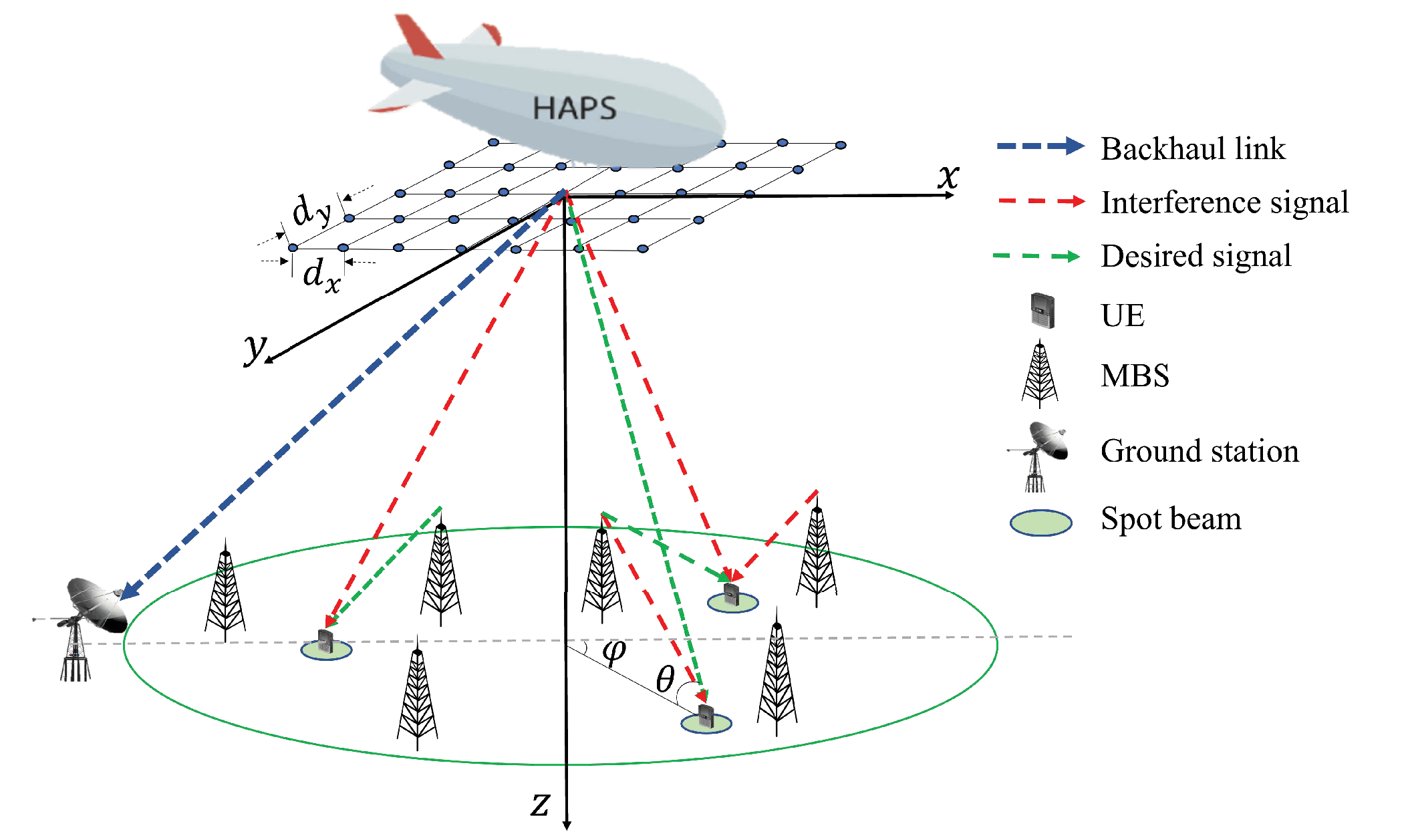}
    \caption{Network architecture.}
    \label{fig_1}
\end{figure}
This paper considers a vHetNet deployed in an urban area, consisting of one HAPS and $B$ MBSs, providing service to $U$ single antenna UEs within a geographically overlapped urban region. The base stations\footnote{In this work, the term base station refers to MBSs and HAPS, collectively.} and UEs are indexed by $b\in\{1,\ldots,B+1\}$ and $u\in \{1,\ldots,U\}$, respectively, with the index $b=B+1$ reserved for HAPS. We assume that the vHetNet operates in the downlink channel and all the UEs share the same time-frequency resource. The considered system operates within the sub-6 GHz band, and each base station $b$ is equipped with atotal number of $N_b=N^\text{V}_b \times N^\text{H}_b,$ antenna elements, where $N^\text{V}_b$ and $N^\text{H}_b$ refer to the number of antenna elements in vertical and horizontal axis, respectively. We refer to each base station's coverage area as a cell.

In this study, as illustrated in Fig.~\ref{fig_1}, each UE is allocated a dedicated spot beam created by appropriate beamforming weights at the serving base station. The beamforming matrix at base station $b$ is represented by $\mathbf{W}^{b}=[\mathbf{w}^b_1,\dots,\mathbf{w}^b_U] \in \mathbb{C}^{N_b \times U}$, where $\mathbf{w}^b_u=[w^b_{1,u},\dots,w^b_{N_b,u}]^T \in \mathbb{C}^{N_b}$, is defined as the beamforming weight vector of UE $u$ at base station $b$. Additionally, we assume that the channel coefficients between each base station $b$ and $U$ UEs form a channel matrix $\mathbf{H}^b=[\mathbf{h}^b_1,\dots,\mathbf{h}^b_U] \in \mathbb{C}^{N_b \times U}$, where the channel vector between base station $b$ and UE $u$ is denoted as $\mathbf{h}^b_u=[h^b_{1,u},\dots,h^b_{N_b,u}]^T \in \mathbb{C}^{N_b}$. We assume that perfect CSI is available at each base station.

Considering the characteristics of urban terrestrial networks where the connections between MBS and UE have dominant NLoS and negligible LoS components, we consider small-scale fading, free-space path loss (FSPL), and shadowing for the channel between each MBS and UE. In this regard, the channel coefficient between antenna element $r \in \{1,\ldots, N_{b}\}$ of MBS $b$, and UE $u$, denoted as $h^{b}_{r,u}$, can be formulated as follows\cite{Terrestrial Channel}:
\begin{equation}\label{MBSchannelgain}
    h^b_{r,u}=\frac{\hat{h}^b_{r,u}\xi^b_u} {\sqrt{{PL}_{b,u}}},~\forall r,~\forall u,~\forall~b \in \{1,\dots,B\},
\end{equation}
where $\hat{h}_{r,u}^b\sim \mathcal{NC}(0,1)$, normal random variable with zero mean and unit variance, represents the small-scale Rayleigh fading channel gain, and $\xi^b_u=10^{{{\xi_u^{'b}}}/10}$ denotes the log-normal shadowing gain, where ${\xi_u^{'b}}$ is the Gaussian random variable with zero mean and standard deviation $\sigma_\xi$ (in dB). ${PL}_{b,u}$ refers to the path loss between base station $b$ and UE $u$, calculated as ${{PL}_{b,u}}=({4\pi f_c d_{b,u}}/{c})^2,~\forall u,~\forall b,$ where $f_c$ is the carrier frequency (in Hz), $d_{b,u}$ (in m) is the distance between the base station $b$ and UE $u$, and $c$ represents the speed of light in free space.

Considering the position of HAPS, located at an altitude 20 Km above the ground, $\mathbf{h}^{B+1}_u$, representing the channel vector between HAPS and UE $u$, can be modeled as a 3D Rician fading channel with a dominant LoS and additional NLoS parts as \cite{HAPS-MIMO}
\begin{equation}
    \mathbf{h}_u^{B+1}=\frac{1}{\sqrt{{PL}_{(B+1),u}}}(\sqrt{\frac{1}{1+K_u}}\hat{\mathbf{h}}_u+\sqrt{\frac{K_u}{1+K_u}}\overline{\mathbf{h}}_u),~\forall u,
\end{equation}
where ${PL}_{(B+1),u}$ represents the FSPL between HAPS and UE $u$, $K_u$ is the Rician factor for UE $u$, and $\hat{\mathbf{h}}_u \in \mathbb{C}^{N_{B+1}}$ represents the NLoS component of the channel vector with its elements from a normal random distribution with zero mean and unit variance, $\mathcal{NC}(0,1)$. Accordingly, $\overline{\mathbf{h}}_u$ is the LoS component of the channel vector as
\begin{equation}\label{LoS-HAPS}
    \overline{\mathbf{h}}_u=\mathbf{a}(\theta_u,\phi_u) \otimes \mathbf{b}(\theta_u,\phi_u),~\forall u,
\end{equation}
where the symbol $\otimes$ denotes the Kronecker product of two vectors, and $\mathbf{a}(\theta_u,\phi_u)$ and $\mathbf{b}(\theta_u,\phi_u)$ are steering vectors defined as
\begin{IEEEeqnarray}{lcl}\label{ab}
    \mathbf{a}(\theta_u,\phi_u)=[1,e^{j2\pi d_h}, \dots, e^{j2\pi (N^{\text{H}}_{B+1}-1)d_h}]^T,~\forall u, \IEEEyesnumber \IEEEyessubnumber* \label{a}\\
    \mathbf{b}(\theta_u,\phi_u)=[1,e^{j2\pi d_v}, \dots, e^{j2\pi (N^{\text{V}}_{B+1}-1)d_v}]^T,~\forall u, \label{b}
\end{IEEEeqnarray}
where $d_h=d_\text{x} \cos{\theta_u}\sin{\phi_u}/\lambda$ and $d_v=d_\text{y} \cos{\theta_u}\cos{\phi_u}/\lambda$. $\theta_u \in [0,\pi/2]$ and $\phi_u \in [-\pi,\pi)$ are the elevation and azimuth angles of UE $u$. $d_\text{x}$ and $d_\text{y}$ are the antenna elements spacing in $x$ and $y$ directions (as shown in Fig.~\ref{fig_1}), and $\lambda=c/f_c$ represents the wavelength.

Now, we formulate the MMF problem. To this end, first, we define the binary user association matrix, $\mathbf{A}\in~\mathbb{B}^{(B+1)\times U}$, whose elements are defined as 
\begin{equation}{\label{eq:User_association}}
    {a_{b,u}} = \begin{cases}
    {1,}&{\text{if UE $u$ is associated to base station $b$}},\\ 
    {0,}&{\text{otherwise.}}
    \end{cases}
\end{equation}

For the considered system model, the received signal of UE $u$ can be expressed as 
\begin{equation}\label{eq:received}
    y_{\text{u}}=\sum_{b=1}^{B+1}{{\left(\mathbf{h}^{b}_{u}\right)^H\mathbf{w}^{b}_{u}}}s_u
    +\sum_{b=1}^{B+1}{\sum_{\substack{k=1\\ k\not= u}}^{U}{{\left(\mathbf{h}^{b}_{u}\right)^H\mathbf{w}^{b}_{k}}s_k}}+n_u,
\end{equation}
where $s_u$ refers to the transmitted data signal of UE $u$. In equation \eqref{eq:received}, the first and second terms refer to the desired and interference signals, received by UE $u$, respectively. Moreover, $n_u$ denotes the additive white Gaussian noise (AWGN) at UE $u$, with zero mean and variance of $\sigma^2_n$. $(\cdot)^H$ refers to the conjugate transpose operation. Henceforth, the signal-to-interference-plus-noise ratio (SINR) for UE $u$ can be defined as
\begin{equation}\label{eq:SINR}
    \gamma_u=\cfrac{\sum_{b=1}^{B+1}{\left|\left(\mathbf{h}^{b}_{u}\right)^H \mathbf{w}^{b}_{u}\right|^2}}{\sum_{b=1}^{B+1}\sum_{\substack{k=1 \\ k\not=u}}^{U}{\left|\left(\mathbf{h}^{b}_{u}\right)^H \mathbf{w}^{b}_{k}\right|}^2+\sigma^2_n},~\forall u,
\end{equation}
where $\mathbf{w}^{b}_{u} \in \mathbb{C}^{N_b}$ represents the beamforming vector for UE $u$ at base station $b$, whereas $\mathbf{h}^{b}_{u} \in \mathbb{C}^{N_b}$ represents the channel vector between base station $b$ and UE $u$. It is noteworthy that the equations \eqref{eq:received} and \eqref{eq:SINR} are applicable under the assumption that the beamforming weights for non-associated UEs are zero at each base station.
Now, we formulate the fairness optimization problem for joint user association and beamforming matrix design. The objective is to maximize the minimum SE of the network, that is equivalent to maximizing the minimum SINR of the network. The initial formulation of the optimization problem can be represented mathematically as
\begin{IEEEeqnarray*}{lcl}\label{eq:P1}
    &\underset{\mathbf{A},~\mathbf{W},{{\gamma}_m}}{\text{maximize}}\,\, & ~~\gamma_m \,  \IEEEyesnumber \IEEEyessubnumber* \label{eq:P1_Obj}\\
    &\text{s.t.} & \gamma_u \geq {\gamma_m},~\forall u, \label{eq:P1_const1}\\
    && \|\mathbf{w}^b_{u}\|^2_2 \leq P^{\text{max}}_b a_{b,u},\,~\forall u,~\forall b, \qquad \label{eq:P1_const2}\\
    && \sum_{u=1}^U a_{(B+1),u} F \log_2(1+\gamma_u) \leq {R_{\text{BH}}}, \qquad \label{eq:P1_const1_BH}\\
    && {\|\mathbf{W}^b \|^2_F}\leq P^{\text{max}}_b, \, ~ \forall b,\qquad \label{eq:P1_const3}\\
    && \sum_{b=1}^{B+1}{a_{b,u}} \geq 1, \,  ~\forall u, \label{eq:P1_const4}\\
    && \mathbf{A} \in \mathbb{B}^{(B+1) \times U}, \label{eq:P1_const5}\\
    && \mathbf{W}^b \in \mathbb{C}^{N_b \times U},~\forall b, \qquad\label{eq:P1_const6}
\end{IEEEeqnarray*}
where the objective function, ${\gamma}_m$, represents minimum SINR of network, and $\mathbf{W}$ denotes the collection of all beamforming matrices $\mathbf{W}^b \in \mathbb{C}^{N_b \times U},~\forall b$. $F$ represents the allocated bandwidth to each UE, $R_{\text{BH}}$ denotes the achievable rate of the backhaul link between HAPS and the ground station, and $P_b^{\text{max}},~\forall b,$ is the maximum available transmit power at base station $b$. $\|\cdot\|_F$ refers to the matrix Frobenius norm operation.

Problem \eqref{eq:P1} aims to jointly design the user association and beamforming matrices, under a set of constraints \eqref{eq:P1_const1}-\eqref{eq:P1_const6}. Constraint \eqref{eq:P1_const1} ensures that the MMF is maintained within the network, while constraint \eqref{eq:P1_const2} enforces that at each base station, the beamforming weights for non-associated UEs are zero. Constraint \eqref{eq:P1_const1_BH} ensures that the sum of the data rate for the UEs, associated with HAPS, does not exceed the achievable rate of the HAPS backhaul link. Constraint \eqref{eq:P1_const3} limits the total transmit power of each base station to its maximum available transmit power. Finally, constraint (\ref{eq:P1_const4}) guarantees that each UE gets associated with at least one base station\footnote{In this study, there is no constraint on the number of associated UEs with each base station.}. Problem \eqref{eq:P1} is an MINLP problem due to non-convex constraints \eqref{eq:P1_const1} and \eqref{eq:P1_const1_BH}, and binary variables in \eqref{eq:P1_const5}. The presence of binary variable matrix $\mathbf{A}$ makes the problem challenging to be solved for the global solution. In the following section, we propose a rapid converging SCA-based\cite{SCA} iterative algorithm to obtain the suboptimal solution for problem \eqref{eq:P1}. The approach utilizes RLT to transform MINLP into an equivalent tractable problem, which is then solved iteratively until convergence to the suboptimal solution.
\section{Problem Relaxation and Proposed Algorithm} \label{Sec:Algorithm}
This section presents a methodology to transform problem (\ref{eq:P1}) into a mathematically tractable form by applying various approximations and techniques to deal with the non-convex parts. These techniques are summarized in the following of this section. 
First, we deal with the non-convex constraint \eqref{eq:P1_const1}. Due to the fractional formula on the left-hand side, the constraint is non-convex. To tackle this, we define a few set of slack variables $\beta_u,~\forall u$, and $\alpha_u,~\forall u$, and replace the left-hand side by its lower bound approximation $\alpha_u$. Two new constraints, as \eqref{betaalpha}, should be defined to ensure that $\alpha_u$ is the lower bound approximation for $\gamma_u$:
\begin{IEEEeqnarray*}{lcl}\label{betaalpha}
    \alpha_u \beta_u \leq \sum_{b=1}^{B+1}{\left|\left(\mathbf{h}^{b}_{u}\right)^H \mathbf{w}^{b}_{u}\right|^2},~\forall u, \IEEEyesnumber \IEEEyessubnumber* \label{eq:alphabetaileqPRi}\\
    \beta_u \geq \sum_{b=1}^{B+1}\sum_{\substack{k=1 \\ k\not=u}}^{U}{\left|\left(\mathbf{h}^{b}_{u}\right)^H \mathbf{w}^{b}_{k}\right|}^2+\sigma^2_n,\,~\forall u. \label{eq:betaigeqIi}
\end{IEEEeqnarray*}

Note that, equation \eqref{eq:betaigeqIi} represents a convex inequality constraint but equation \eqref{eq:alphabetaileqPRi} is non-convex due to the norm at the right hand side. To handle this, we define slack variable matrices $\mathbf{P} \in \mathbb{R}^{(B+1) \times U},$ and $\mathbf{Q} \in \mathbb{R}^{(B+1) \times U},$ with their elements, respectively, defined as
\begin{IEEEeqnarray*}{lcl}\label{Real and Imag}
    p_{b,u} \leq \Re{\left(\left(\mathbf{h}^{b}_{u}\right)^H \mathbf{w}^{b}_{u}\right)},~\forall u,~\forall b, \qquad \IEEEyesnumber \IEEEyessubnumber* \label{Real}\\
    q_{b,u} \leq \Im{\left(\left(\mathbf{h}^{b}_{u}\right)^H \mathbf{w}^{b}_{u}\right)},~\forall u,~\forall b, \qquad\label{Imag}
\end{IEEEeqnarray*}
where, $\Re (\cdot)$ and $\Im (\cdot)$ represent the real and imaginary parts of a complex number, respectively.
Consequently, equation \eqref{eq:alphabetaileqPRi} can be reformulated as 
\begin{equation}\label{eq:alphaleqV2}
    \alpha_u \beta_u \leq {\sum_{b=1}^{B+1}{(p_{b,u})^2+(q_{b,u})^2}},~\forall u.
\end{equation}

In order to be able to convert equation \eqref{eq:alphaleqV2} to a convex form, we rewrite the equation as
\begin{equation}\label{betaalpha2}
    \alpha_u \leq \sum_{b=1}^{B+1}{\frac{(p_{b,u})^2+(q_{b,u})^2}{\beta_u}},~\forall u,
\end{equation}
which can be converted to equivalent convex form by replacing the right hand side with its approximated first-order Taylor series as
\begin{multline}{\label{first Taylor}}
     \alpha_u \leq \sum_{b=1}^{B+1}\frac{2{p^{(n)}_{b,u}}}{{\beta^{(n)}_{u}}}\left(p_{b,u}-{p^{(n)}_{b,u}}\right)+\frac{2{q^{(n)}_{b,u}}}{{\beta^{(n)}_u}}\left(q_{b,u}-{q^{(n)}_{b,u}}\right)\\ +\frac{\left({p^{(n)}_{b,u}}\right)^2+\left({q^{(n)}_{b,u}}\right)^2}{\beta^{(n)}_u}\left(1-\frac{\beta_u-{\beta^{(n)}_u}}{{\beta^{(n)}_u}}\right),~\forall u,
\end{multline}
where superscript $(n)$ refers to the value of corresponding variable in the $n$th iteration of the SCA process. Next, we tackle constraint \eqref{eq:P1_const1_BH} by acknowledging that at the convergence, $\gamma_u, \forall u,$ will converge to $\gamma_m$\cite{MMF_Convergence}, and by employing Jensen's inequality\footnote{$\mathbb{E}\{\log(f(x))\} \leq \log(\mathbb{E}\{f(x)\})$, where $\mathbb{E}$ denotes the expected value.}. Consequently, we can replace constraint \eqref{eq:P1_const1_BH} as
\begin{equation}\label{BH_Constraint_V2}
    \sum_{u=1}^U a_{(B+1),u}\gamma_m \leq {2^{R_{\text{BH}}/{F}}-1}.
\end{equation}

Furthermore, we define slack variables $\eta_u=a_{(B+1),u} \gamma_m,\forall u,$ and define four new constraints as below to ensure that the actual equality holds:
\begin{IEEEeqnarray*}{lcl}\label{eq:BH_1}
    0\leq \eta_u \leq \gamma_{\text{max}} a_{(B+1),u},~\forall u,\IEEEyesnumber \IEEEyessubnumber* \label{new_a}\\
    0\leq \gamma_m-\eta_u \leq \gamma_{\text{max}}(1-a_{(B+1),u}),~\forall u,\label{new_b}
\end{IEEEeqnarray*}
where $\gamma_{\text{max}}$ is enough large number denoting the upper bound for $\gamma_m$. Consequently, constraint \eqref{BH_Constraint_V2} will be transformed as
\begin{equation}\label{eq:BH_2}
    \sum_{u=1}^U \eta_u \leq {2^{R_{\text{BH}}/{F}}-1}.
\end{equation}
\vspace{1mm}
After applying the aforementioned relaxations, the SCA optimization problem at $n$th iteration will be as
\begin{IEEEeqnarray}{lcl}\label{eq:P2}
    &\underset{\mathbf{W},\mathbf{A},\boldsymbol{\alpha},\boldsymbol{\beta},\mathbf{P},\mathbf{Q},\boldsymbol{\eta},\gamma_m}{\text{maximize}}\,\, & \gamma_m \,  \IEEEyesnumber \IEEEyessubnumber* \label{eq:P2_Obj}\\
    &\text{s.t.} & \alpha_u \geq \gamma_m,~\forall u,\,\\
    && \eqref{eq:P1_const2},\eqref{eq:P1_const3},\eqref{eq:P1_const4},\eqref{eq:P1_const5},\eqref{eq:P1_const6},\\
    && \eqref{eq:betaigeqIi},\eqref{Real and Imag},\eqref{first Taylor},\eqref{eq:BH_1},\eqref{eq:BH_2},\qquad 
\end{IEEEeqnarray}
where $\boldsymbol{\alpha}=[\alpha_1,\dots,\alpha_U]^T$, $\boldsymbol{\beta}=[\beta_1,\dots,\beta_U]^T$, and $\boldsymbol{\eta}=[\eta_1,\dots,\eta_U]^T$. Problem \eqref{eq:P2} is a second order convex programming (SOCP) type problem which can be solved using solvers such as Mosek. The pseudocode of the proposed iterative joint user association and beamforming weights design (JUBD) algorithm is outlined in Algorithm~\ref{alg:alg1}. According to the algorithm, \eqref{eq:P2} is solved iteratively until the difference ratio of the objective function value is less than the tolerance $10^{-4}$, or maximum number of SCA iterations, $N_{\text{iter}}$, is reached, whichever first. Since the objective function of problem \eqref{eq:P2} is bounded above due to the maximum transmit power constraint \eqref{eq:P1_const3}, and each iteration of Algorithm~\ref{alg:alg1} generates a sequence of improved objective function values, Algorithm~\ref{alg:alg1} converges to a stationary point of problem \eqref{eq:P1}\cite{SCA}. In Algorithm~\ref{alg:alg1}, the initial values are chosen from the feasible set. The worst-case computational cost of the algorithm is mainly determined by the branch-and-bound method to solve MIP at each SCA iteration, and is given as $\mathcal{O}(2^{U\left(B+1\right)})$.
\begin{algorithm}[h!]
\caption{Proposed iterative JUBD algorithm for vHetNets.}\label{alg:alg1}
    \begin{algorithmic}[1]
    \small
    \STATE \textbf{Input:}~$U,~B,~N_{b},~\forall b,~\mathbf{H}^b,~\forall b,~P_b^{\text{max}},~\forall b,~\sigma^2_n,~F,~R_\text{BH},~N_{\text{iter}}=10$.\\
    \STATE \textbf{Output:} $\mathbf{W}^*,~\mathbf{A}^*$.\\
    \STATE \text{Initialize} $\boldsymbol{\beta}^{(0)},~\mathbf{P}^{(0)},~\mathbf{Q}^{(0)},$ \text{and set} $n:=0.$\\
    \STATE \text{Repeat until $|\gamma_m^{(n)}-\gamma_m^{(n-1)}|/\gamma_m^{(n-1)} \leq 10^{-4} $} or until $n \leq N_{\text{iter}}:$\\
    \hspace{0.001cm} \text{- solve (\ref{eq:P2}) to find $~\mathbf{A}^{(n)*},~\mathbf{W}^{(n)*}, ~\boldsymbol{\alpha}^{(n)*}, ~\boldsymbol{\beta}^{(n)*},~\mathbf{Q}^{(n)*},$}\\
    \text{$\mathbf{P}^{(n)*},~\boldsymbol{\gamma}_m^{(n)*}$}.\\
    \hspace{0.001cm} \text{- update $\boldsymbol{\beta}^{(n+1)}:=\boldsymbol{\beta}^{(n)*},~\mathbf{P}^{(n+1)}:=\mathbf{P}^{(n)*},$}\\
    \text{$\mathbf{Q}^{(n+1)}:=\mathbf{Q}^{(n)*},~n:=n+1$}.
    \end{algorithmic}
\label{alg1}
\end{algorithm}
\section{Numerical Results} \label{Sec:Results}
 \begin{table*}[t]
\caption{Simulation Parameters.}\label{tab:table1}
\centering
\begin{tabular}{|c||c|}
\hline
\textbf{Parameter} & \textbf{Value}\\
\hline
Center frequency ($f_c$) & $2.545$ GHz\\
\hline
Shadowing standard deviation ($\sigma_\xi$), Rician factor ($K_u$) & $8$,~$10$\\
\hline
Number of antenna elements ($N_{\text{b}},~b\in \{1,\dots,B\}$,~$N_{B+1}$) & $4 \times 4$,~$8 \times 8$\\
\hline
HAPS antenna elements spacing ($d_x,~d_y$) & $\lambda/2$\\
\hline
Maximum transmit power ($P_b^{\text{max}},~b\in \{1,\dots,B\}$,~$P_{B+1}^{\text{max}}$) & $43$ dBm,~$52$ dBm\\
\hline
UE allocated bandwidth ($F$), HAPS backhaul rate ($R_\text{BH}$) & $1$ MHz,~$20$ Gbps\\
\hline
AWGN variance ($\sigma^2_n$) & $-100$ dBm\\
\hline
\end{tabular}
\end{table*}
\begin{figure}[!t]
        \centering
        \includegraphics[width=\linewidth]{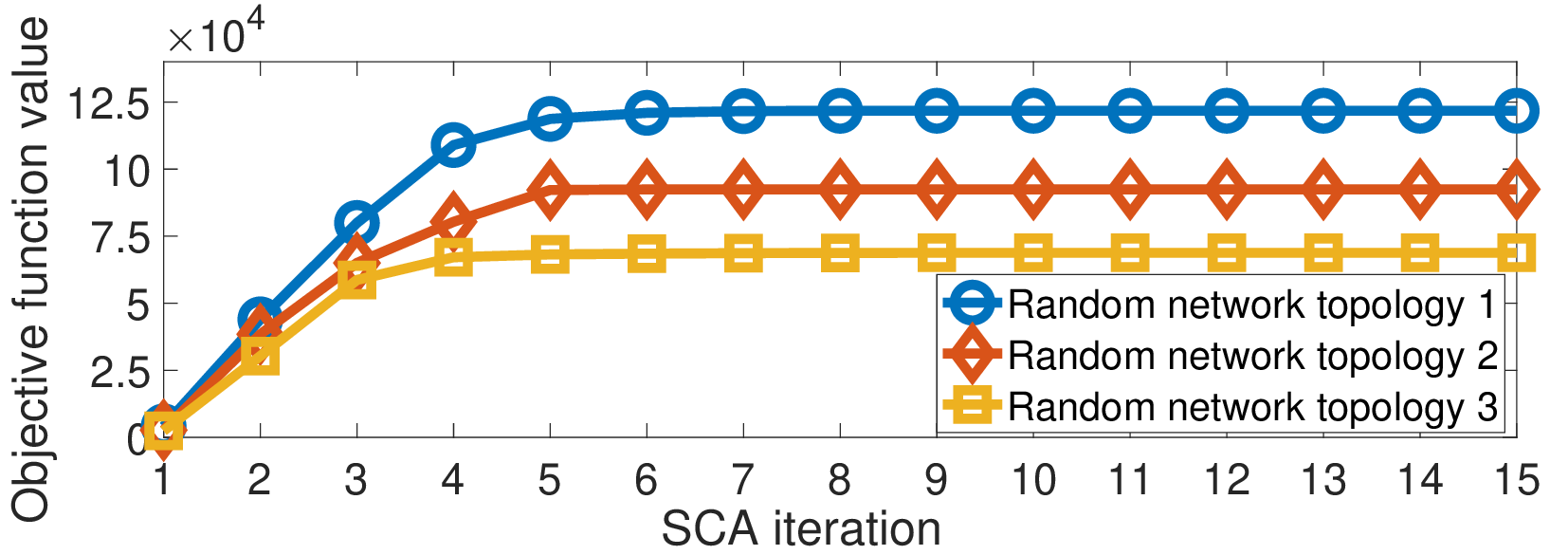}
    \vspace{-3mm}
    \caption{Convergence behaviour of the proposed Algorithm~\ref{alg:alg1}.}
    \label{fig_3}
\end{figure}
\begin{figure*}[t]
    \centering
    \captionsetup{justification=centering}
    \begin{subfigure}{\columnwidth}
        \centering
        \includegraphics[width=\columnwidth]{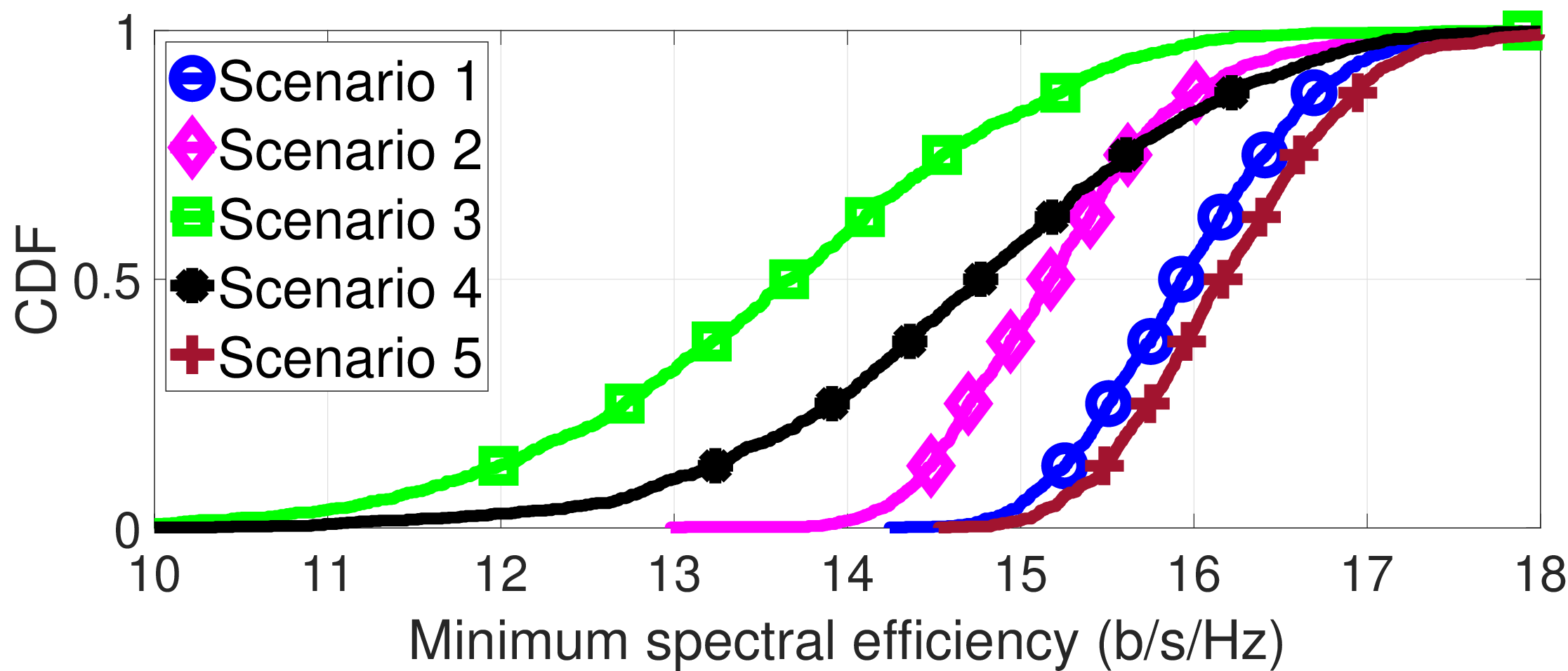}
        \caption{CDF of minimum spectral efficiency for 5 Scenarios.}
        \label{fig_4:a}
    \end{subfigure}
    \begin{subfigure}{\columnwidth}
        \centering
        \includegraphics[width=\columnwidth]{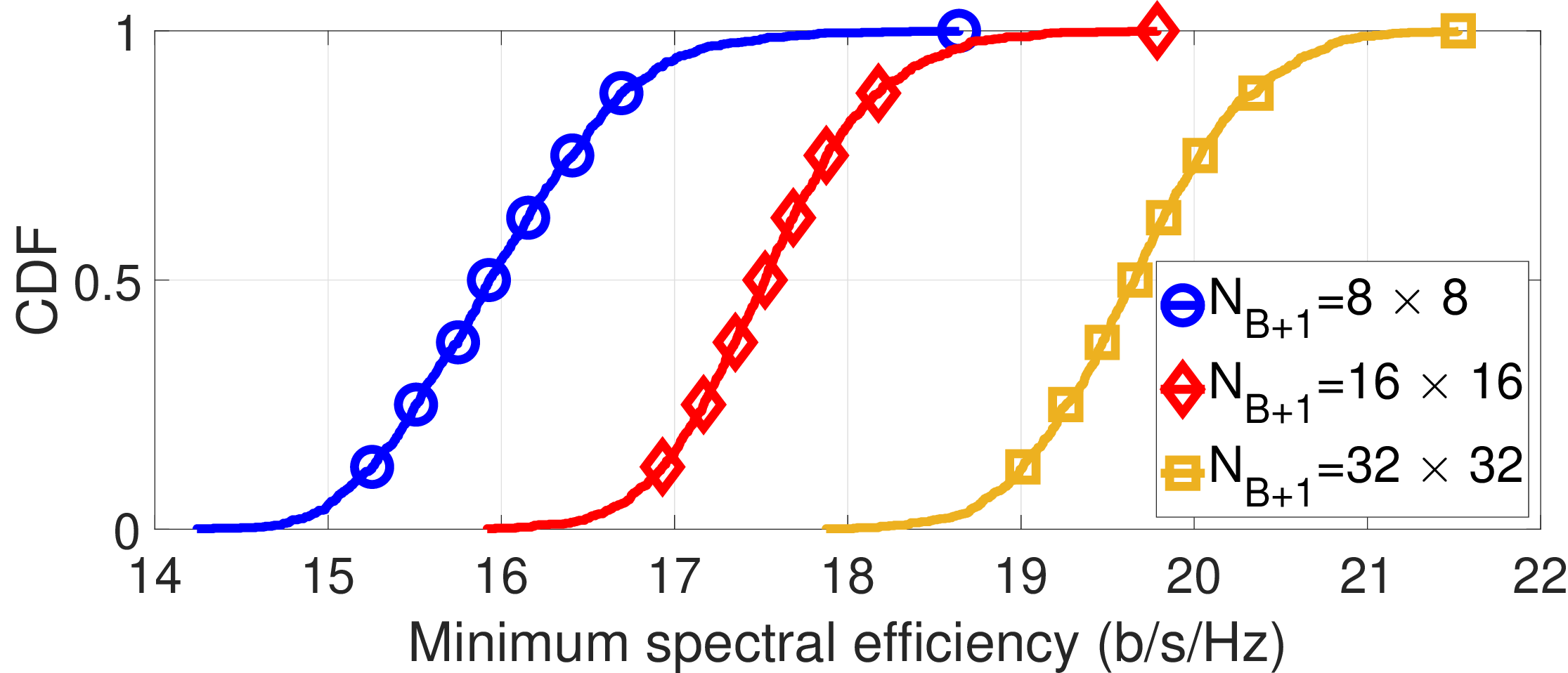}
        \caption{Impact of HAPS~antenna~configuration in Scenario 1.}
        \label{fig_4:b}
    \end{subfigure}
    \vspace{-1mm}
    \caption{\small Statistical~behavior~of~minimum~spectral~efficiency.~Scenario~1:~vHetNet~(4~MBSs~+~HAPS)~using~JUBD~Algorithm. Scenario~2:~vHetNet~(4~MBSs~+~HAPS)~beamforming~optimization~with~max-SINR~based~user~association.~Scenario~3:~Standalone terrestrial network (4 MBSs) using JUBD Algorithm. Scenario~4:~Standalone~terrestrial~network~(5~MBSs)~using~JUBD~Algorithm. Scenario~5:~vHetNet~(4~MBSs~+~HAPS)~using~JUBD~Algorithm with multiple base station association.}
    \label{fig_4}
\end{figure*}
\begin{figure}
    \centering
    \includegraphics[width=\linewidth]{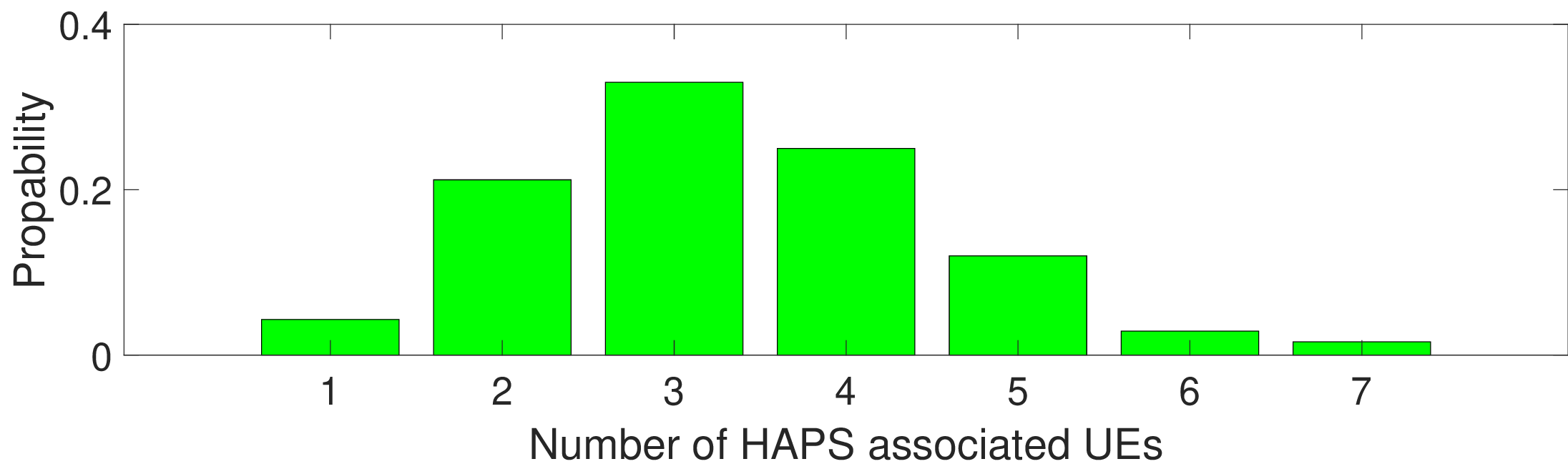}
    \vspace{-5mm}
    \caption{\small Probability distribution of UEs associated with HAPS.}
    \label{fig:5}
\end{figure}
In this section, we evaluate the performance of the proposed JUBD Algorithm~\ref{alg:alg1} through numerical simulations. In simulations, we consider a $4$ Km by $4$ Km square hotspot urban area with $16$ uniformly distributed UEs in it. To facilitate comparisons, we consider five distinct scenarios. Scenario 1 involves a vHetNet comprising four MBSs and one HAPS, and employs JUBD algorithm~\ref{alg:alg1}. Scenario 2 also involves a vHetNet comprising four MBSs and one HAPS but focuses only on optimizing the beamforming weights and employs the maximum received SINR-based scheme for user association. Scenarios 3 and 4 consider a standalone terrestrial network comprising four and five MBSs, respectively, and employ JUBD Algorithm~\ref{alg:alg1}. Scenarios 1 to 4 assume that each UE can get associated to only one base station; however, we consider Scenario 5 where Algorithm~\ref{alg:alg1} is implemented in a vHetNet (one HAPS and four MBSs) with multiple base station association, where UEs can get associated with more than one base station. Another potential scheme to compare with our proposed JUBD Algorithm 1 is the DQL method\cite{UA} which may not be the best for the problem setting in this paper due to the larger scale of the network and a larger number of variables. The results are obtained for each scenario based on $1000$ independent and identically distributed (i.i.d.) network topologies. All simulations are implemented in MATLAB using CVX with Mosek 9.1.9 as the internal solver. The rest of the simulation parameters are listed in Table~\ref{tab:table1}.

Fig.~\ref{fig_3} illustrates the evolution of objective function values across SCA iterations for three randomly sampled network topologies representing UE distributions. It can be observed that the objective function value increases with each SCA iteration and saturates after five or six iterations.
Fig.~\ref{fig_4} illustrates the cumulative distribution function (CDF) of the minimum SE for different scenarios. Particularly, Fig.~\ref{fig_4:a} compares Scenarios 1, 2, 3, 4, and 5, and from it, we can deduce four observations. First, the proposed vHetNet architecture (i.e., Scenarios 1, 2, and 5), provides improved SE performance as compared to those of standalone terrestrial networks (i.e., Scenarios 3 and 4). This improvement is due to the fact that vHetNets allow some UEs to connect to HAPS with higher order MIMO antenna, LoS links and better channels, resulting in improved SE performance. Fig.~\ref{fig:5}, which plots the probability distribution of UEs associated with HAPS, corroborates this fact that in every sample network topology, a few number of UEs (less than seven UEs out of 16) are associated with HAPS.

Secondly, the proposed JUBD Algorithm~\ref{alg:alg1} for vHetNet, used in Scenarios 1 and 5, yields better SE performance compared to the scenario where only the beamforming weights are optimized (i.e., Scenario 2). Thirdly, the CDF curve of vHetNet (i.e., Scenarios 1, 2, and 5) exhibited a notably steeper slope when compared to those of standalone terrestrial networks (i.e., Scenarios 3 and 4). Steepness indicates a reduced standard deviation concerning the minimum SE. Accordingly, in vHetNets, UEs experience almost the same quality-of-service (QoS) regardless of their positioning within the coverage area. This is because if a UE experiences suboptimal channel conditions within the terrestrial network, it gets associated with HAPS with a better channel quality due to the dominant LoS connection. The third observation reaffirms the provision of fairer QoS to UEs in different network topologies. Lastly, comparing Scenarios 1 and 5, it can be observed that Scenario 5, where each UE can get associated with multiple base stations in a vHetNet, leads to a slightly higher SE compared to Scenario 1, where UEs can get associated with only one base station.

Fig.~\ref{fig_4:b} plots the CDF of minimum SE for Scenario 1 with the varying number of antenna elements on HAPS, $N_{B+1}=\{8 \times 8,~16 \times 16,~32\times 32\}$. It can be observed that the SE performance of the network increases with the number of antenna elements on HAPS. This is because a higher order MIMO antenna easily forms a narrower spot beam, resulting in increased gain toward the target UE and decreased interference toward other UEs.
\section{Conclusion} \label{Sec:Conclusion}
In this work, we proposed the design of an optimum beamforming weight and user association scheme for an integrated HAPS-terrestrial network (vHetNet) to mitigate interference. We formulated an MMF optimization problem, which turned out to be a MINLP. In general, MINLP problems are challenging to solve; thus, we reformulated the original problem and developed a low-complexity and fast-converging SCA-based iterative algorithm to solve it. Simulation results showed that vHetNets with optimized beamforming and user association scheme offer higher SE than standalone terrestrial network scenarios. This observation demonstrates the potential of HAPS-enabled vHetNets for the 6G and beyond wireless networks. The interference management in vHetNets, based on machine learning approaches, opens up future research directions.

\end{document}